\documentclass[journal,twoside,web]{ieeecolor}

\usepackage{generic}
\usepackage{cite}
\usepackage{amsmath,amssymb,amsfonts}
\usepackage{algorithmic}
\usepackage{graphicx}
\usepackage{textcomp}
\usepackage{caption}
\usepackage{multirow}
\usepackage{array}
\usepackage{booktabs}
\usepackage{subfigure}

\def\BibTeX{{\rm B\kern-.05em{\sc i\kern-.025em b}\kern-.08em
    T\kern-.1667em\lower.7ex\hbox{E}\kern-.125emX}}
\begin{document}
\title{Remote blood pressure measurement via spatiotemporal mapping of a short-time facial video}
\author{Jialiang Zhuang, Bin Li, Yun Zhang, Yuheng Chen, Xiujuan Zheng
\thanks{Jialiang Zhuang, yuheng Zhang and Xiujuan Zheng, are all with College of Electrical Engineering, Sichuan University, Chengdu, Sichuan, China. (Corresponding author: Xiujuan Zheng, xiujuanzheng@scu.edu.cn )}
\thanks{Bin Li, is with School of Computer Science, Northwest University, Xi’an, China}
\thanks{Yun Zhang, is with School of Information Science and Technology, Xi'an Jiaotong University, Xi’an, China}
}

\maketitle
\begin{abstract}
Blood pressure (BP) monitoring is vital in daily healthcare, especially for cardiovascular diseases. However, BP values are mainly acquired through the contact sensing method, which is inconvenient and unfriendly to continuous BP measurement. Hence, we propose an efficient end-to-end network to estimate the BP values from a facial video to achieve remote BP measurement in daily life. In this study, we first derived a Spatial-temporal map of a short-time (~15s) facial video. According to the Spatial-temporal map, we then regressed the BP ranges by a designed blood pressure classifier and simultaneously calculated the specific value by a blood pressure calculator in each BP range. In addition, we also developed an innovative oversampling training strategy to handle the unbalanced data distribution problem. Finally, we trained the proposed network on a private dataset ASPD and tested it on the popular dataset MMSE-HR. As a result, the proposed network achieved a state-of-the-art MAE of 12.35 mmHg and 9.5 mmHg on systolic and diastolic BP measurements, which is better than the recent works. It concludes that the proposed method has excellent potential for camera-based BP monitoring in real-world scenarios.

\end{abstract}


\section{Introduction}
Blood pressure is a primary physiological parameter of the human body and an essential basis for disease diagnosis. Cardiovascular diseases are the leading cause of death globally. Most of them are highly related with hypertension. In this context, convenient ordinary monitoring of blood pressure in daily life becomes critical.\par

Many outstanding scientists\cite{45Daniel2021,46Anand2020} have devoted themselves to the study of calculating blood pressure by extracting the morphology and temporal characteristics of remote photoplethysmographic (rPPG) waveform and obtained good results in the end, which proves that a single pulse wave can effectively reflect the change of blood flow in the human cardiovascular system in different heart cycles. However, most of them need a special contact photoelectric sensor to obtain blood volume pulse (BVP), which greatly limits the application range. Recently many excellent works were proposed that can robustly simulate BVP based on face videos and able to accurately calculate heart rate and heart rate variability. In addition, some works such as RthymNet~\cite{36Niu2019} showed that spatio-temporal map of face videos is capable of exploiting blood flow state in blood vessels, Cheng~\cite{40Cheng2021} also designed a method to convert BVP into pressure pulse wave, indicating that PPG waveform can be filtered from the BP waveform at the same artery and comprise BP information. These results of the previous results enable us to believe that the spatio-temporal map of single point face video can calculate accurate blood pressure value. 
 
 Still, there are many challenges in non-contact blood pressure measurements; the accuracy of such tasks is easily affected by many factors~\cite{2Markandu2000TheMS}, including the professional abilities of the conductors, the height, weight, and age of the subjects, and the seasons and environments. For example, head motion creates difficulties for the region cutting algorithm, causes residual motion image, and degrades the image quality. In addition, the weak remote BVP signal reflecting changes in hemoglobin concentration in the circulatory system can easily be contaminated and submerged by changing ambient light. Moreover, because of difficulties of blood pressure data collection and the privacy issues of face video, there is very few large-scale blood pressure database with face videos. The lack of data greatly hinders the rapid development in the field of non-contact measurement of blood pressure.\par
Therefore, this study is proposing the first end-to-end non-contact blood pressure measurement network based on face videos in this study. The main contributions are summarized as follows:
\begin{enumerate}
    \item Design a continuous BP estimation method only based on spatial-temporal map of facial videos.
    \item Propose a blood pressure classifier to transform a regression problem into a joint problem of classification and regression.
    \item Exploit an oversampling training scheme for blood pressure monitoring task, which effectively addresses the uneven distribution of blood pressure data set in the training process.
    \item Only require a short-time (15 seconds) facial video to measure diastolic and systolic blood pressure.
\end{enumerate}

\subsection{Related work}
Until now, two categories of methods have been used to measure blood pressure: contact and non-contact measurements. Contact blood pressure measurements are made with medical equipment to directly monitor human biological signals~\cite{8Young1990,7KK2021,9Tholl2004,5Beevers2001}, including two main technical approaches. One route is to measure the pressure change with a pressure transducer, for example, arterial manometry~\cite{14Pressman1963}.The other technical approach is to control the pressure applied to the measured area by an external device and then map Coriolis sound realted information to BP values via experience-based correlation. The most classic apparatus is the mercury sphygmomanometer~\cite{5Beevers2001}, and its accuracy largely depends on the professional level of the user.

Non-contact blood pressure measurements usually calculate the systolic and diastolic blood pressure based on the features of BVP signals~\cite{3TK2001}. Pulse wave velocity (PWV) referred as the pressure wave velocity propagates along the wall of the great artery during each cardiac ejection. It is a simple, effective, and economic index for the non-contact evaluation of arterial stiffness~\cite{16B1976,29Mousavi2018}. Moreover, it is an independent predictor of cardiovascular events closely related to blood pressure~\cite{18ML1957,17RA2006,19,41Fan2020}. A series of simulation experiments on humans and animals demonstrate that pulse wave transit time (PWT) is closely related to diastolic and systolic blood pressures. Owing to this theory, a novel indirect method, which differed from the traditional direct methods using cuff or oscilloscope devices, was proposed to measure blood pressure values. In 2014, Rohan et al. introduced that the arterial blood pressure value can be calculated accurately based on the extracted morphology features of BVP signals~\cite{20Samria2014}, which was obatained by the skin-contact noninvasive sensors derived from optical plethysmography (PPG) technology. After detecting the blood volume changes of the cardiac cycle with infrared light, the time delay of systolic and diastolic peak values were also extracted to model the relationship between these features and BP values. Another study added 1/2 pulse width and 2/3 pulse width as the characteristics of BVP signal together with systolic upstroke time and diastolic time to facilitate the model for calculating BP values~\cite{23Teng2003}. As the correlation between blood pressure and BVP signals was assumed as nonlinear and vulnerable to noise, artificial neural networks, which is good at feature extraction, were adopted to achieve satisfactory results of estimated BP values.~\cite{27Kurylyak2013,24Wang2018}.\par

In order to extract as much information as possible, and then investigate the best combination, some study considered additional data source.Some researchers~\cite{21Shen2015} proposed that electrodes and instruments could collect ECG signals while PPG signals were obtained. Others designed a new sensor to integrate PPG signal and ECG signal simultaneously~\cite{22Han2020} and then eliminated baseline drift noise by wavelet change, which significantly improved the measured results. Similar to the development route of blood pressure detection method with only PPG signal, some papers~\cite{26Jung2005} also used artificial neural network (ANN) to train the network model to predict blood pressure according to the sample experiment~\cite{26Jung2005,25Maher2020}. Based on previous work, demographic information (including weight, height, and other factors) was added as the input of network for training. The results showed that the performance of the mode~\cite{44Rong2021}l was much better than that of the multiple regression method. In addition to the use of artificial neural networks and other network models, traditional machine learning method~\cite{29Mousavi2018,30He2016,28Zhang2017} was also expolited in many studies to optimize the feature extraction process. In 2021 Meng et al.~\cite{44Rong2021} showd that temporal and energy features of BVP signal obtained by face videos can also be adopted to estimate blood pressure comprehensively.\par

\section{Methodology}
\begin{figure}[ht]
	\centering
        		\includegraphics[width=\linewidth]{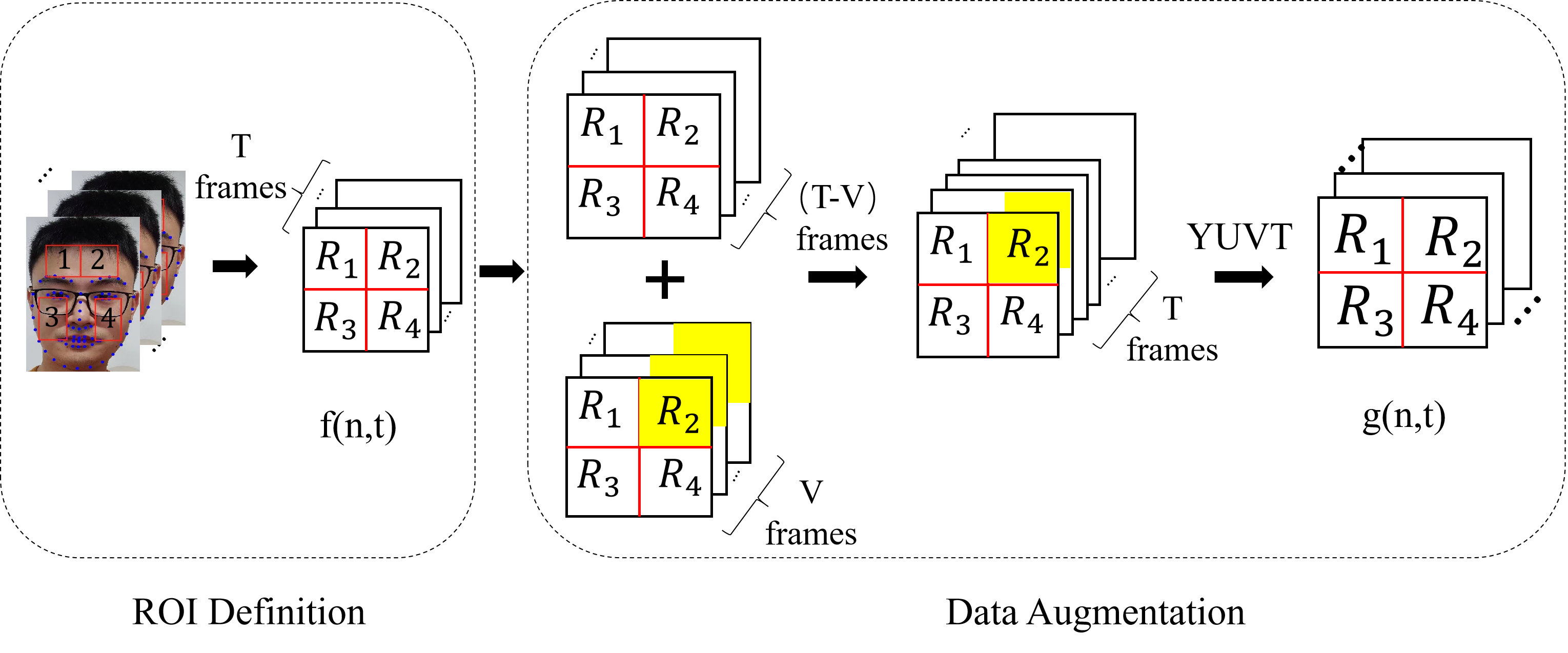}
	  \caption{Overview of the module of ROI Definition and Data Augmentation. Seetaface6 is used to detect human face and localize 68 landmarks, which are then fed into the Data augmentation module to get spatial-temporal map (STM).}
	  \label{fig1}
\end{figure}

This study proposes a non-contact blood pressure measurement network.The overall workflow is shown in Figure~\ref{fig1}. Firstly, four ROIs, which contains the blood pressure signal from the input facial video, was defined based on landmarks detected by SeetaFace6. Secondly, the data augmentation (DA) module randomly mask a part of above cropped video sequence along both time-domain and spatial-domain and transform RGB color space into YUVT color space, aiming to reduce interferences caused by illumination variations and sudden non-rigid motions. Then, the cropped video sequence is sliced to generate the multi-domain spatial-temporal mapping, facilitating the succeeding to map the relationship between video sequence and BP values. Finally, the Blood Pressure Estimator module, including blood pressure classifier and blood pressure calculator, estimates systolic and diastolic blood pressure, respectively. Details of each module are explained in the following subsections.\par

\subsection{ROI Definition}
In order to make full use of informative parts containing efective signals and eliminate the noisy parts such as eyes, nose, hair, and mouth, four regions of interest (ROIs) are defined by selecting the forehead and cheek with rich vascular information. First, the authors use SeetaFace6 to detect face and localize 68 landmarks and select 7 points to delineate the ROIs in each frame in the video sequence, as shown in Figure~\ref{fig1}. Next, we calculate the average pixel values of each ROI to represent physiological information, noted as initial spatial-temporal map (ISTM): \begin{equation}
f(n, t)=\frac{\sum_{\mathrm{x}, \mathrm{y} \in \mathrm{R(n)}} V(x, y, t)}{\mathrm{A(n)}}
\end{equation}
Where R(n) represents the $n_{th}$ ROI, A(n) represents the number of pixels in the selected region, V(x,y,t) represents the pixel value of the $n_{th}$ ROI position (x, y) of frame t of the face video sequence.


\subsection{Data Augmentation}
\begin{figure}[ht]
	\centering
		\includegraphics[width=0.75\linewidth]{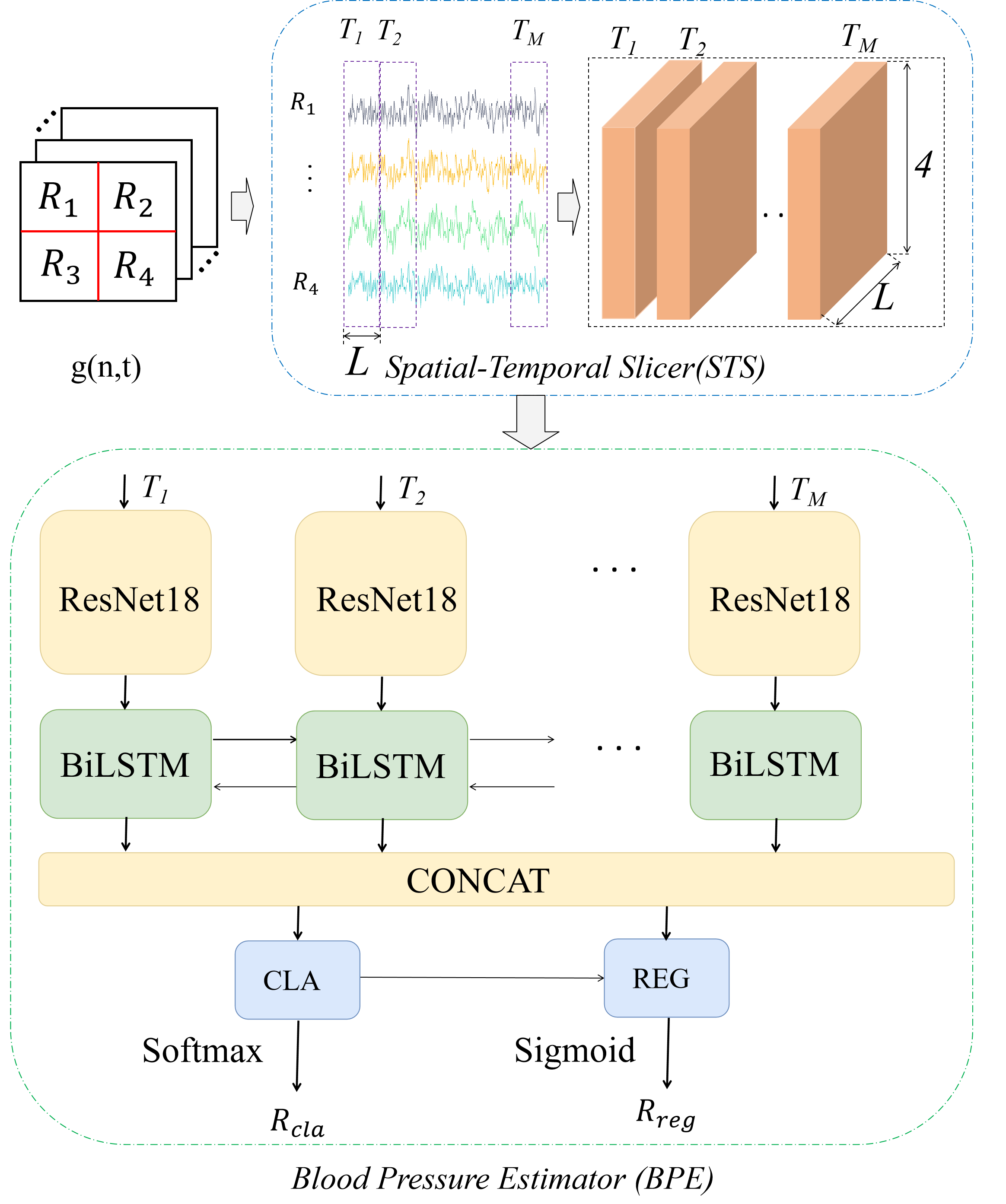}
	  \caption{Architecture of the BPM-Net. Firstly, STM is transformed into multi-domain spatial-temporal feature mapping (MDSTM) $\left\{\mathrm{T_{1}}, \mathrm{T_{2}}, \ldots, \mathrm{T_{M}}\right\}$. Then the feature extractor composed of a series backbone network and BiLSTM fits the high-dimensional feature, which is fed into blood pressure classifier to locates the blood pressure interval through this feature. Finally, the blood pressure calculator combines the results of the feature extractor and the blood pressure classifier to output the final blood pressure value.}
	  \label{fig2}
\end{figure}

Due to the moving artifacts and environmental illumination variation during signal acquisition, the accuracy of non-contact blood pressure detection is easily affected. Therefore, a data augmentation strategy is proposed to solve the problem of signal contamination. First, we randomly mask a small part of the initial spatial-temporal map $f(n,t)$ along both the time dimension and spatial dimension as shown in  Figure~\ref{fig2}. Then, YUV Transformation (YUVT) color space, which gives more attention to brightness dimension features in color space, is proposed here to extract information related to BVP to get spatial-temporal maps (STM) $g\left(n,t\right)$. The new color space transformation can be formulated as:
\begin{small}
 \begin{footnotesize}  
\begin{equation}
\left[\begin{array}{l}
Y_{t}(x, y) \\
U_{t}(x, y) \\
V_{t}(x, y)
\end{array}\right]=\left[\begin{array}{ccc}
0.299 & 0.587 & 0.114 \\
-0.169 & -0.331 & 0.5 \\
0.5 & -0.419 & -0.081
\end{array}\right]\left[\begin{array}{c}
R_{t}(x, y) \\
G_{t}(x, y) \\
B_{t}(x, y)
\end{array}\right]
\label{eq1}
\end{equation}
\end{footnotesize}
\end{small}

where t indicates video frame index, $R_{t}(x, y)$, $G_{t}(x, y)$, $B_{t}(x, y)\in g_{roi}^{\prime}\left(t, r_{i}\right)$, and $R_{t}(x, y)$,$ G_{t}(x, y)$, $B_{t}(x, y)$ represents the pixel value in RGB color space, while $Y_{t}(x, y)$, $U_{t}(x, y)$, $V_{t}(x, y)$ are the pixel value of the YUVT color channel. All of them constitute the video sequence after data enhancement together. The enhanced video sequence is noted as a spatial-temporal map (STM) $g\left(n,t\right)$.\par

\subsection{Architecture}
\subsubsection{Spatial-Temporal Slicer}
The average value of each ROI of STM is flattened and spliced to form the feature vector corresponding to the t frame, which can be written as :
\begin{equation}
\mathrm{SS}(\mathrm{t})=\mathrm{CAT}\left(\emptyset\left(\bigcup_{\mathrm{n}=1}^{\mathrm{N}} g(\mathrm{n}, \mathrm{t})\right)\right)
\end{equation}W
where $\emptyset$ represents the extracting non empty subsets operation, and $\mathrm{CAT}(\cdot)$ represents the feature vector splicing operation.
We combine the eigenvectors of all frames into a spatial-temporal feature map, and divide them into two parts by sliding windows with step length L. M separate time-space fragments $\left\{\mathrm{T_{1}}, \mathrm{T_{2}}, \ldots, \mathrm{T_{M}}\right\}$ are constructed as multi-domain spatial-temporal feature mapping $\mathrm{STS}(m)$, which can be defined as :
\begin{equation}
\mathrm{STS}(m)=\bigcup_{t=N \times c l}^{(N+1) \times c l} \mathrm{SS}(t)
\end{equation}
where m is the index of spatial-temporal slicer. In order to use video information effectively and help network learning, I normalized the pixel values of all video frames to the range of [0,1].

\subsubsection{Blood Pressure Estimator}
In this experiment, the residual convolution neural network~\cite{37Yu2019} is used as the backbone network, and the size of all convolution kernels is 3 × 3. The step is 1. Firstly, the feature extraction network module is used to perform high-dimensional feature extraction on the multi-temporal spatial feature map STS(m). The feature extraction network corresponding to each time domain segment adopts the same parameters. The results are fed to the subsequent LSTM to strengthen the temporal correlation and obtain the high-dimensional semantic features F. Then blood pressure classifier is used to classify the high-dimensional semantic features F to obtain the blood pressure interval. This operation can establish a benchmark for the features and effectively reduce the over fitting phenomenon of the network. Finally, The results of the classifier are integrated with the output of the feature extractor to calculate the specific value of blood pressure, which can also effectively help in classifier training.

We express the whole BPE process as the following formula:
\begin{equation}
\mathrm{F}=\mathrm{CAT}\left(\bigcup\mathrm{FE}\left(\mathrm{STS}(\mathrm{m})\right)\right)
\end{equation}
\begin{equation}
\mathrm{F}_{\mathrm{cla}}=\mathrm{CLA}(\mathrm{F})
\end{equation}

\begin{equation}
\mathrm{R}_{\mathrm{reg}}=\mathrm{REG}\left(\mathrm{CAT}\left(\mathrm{F}, \mathrm{F}_{\mathrm{cla}}\right)\right)
\end{equation}

\begin{equation}
\mathrm{R}_{\mathrm {cla }}=\mathrm{SOFTMAX}\left(\mathrm{F}_{\mathrm{cla}}\right)
\end{equation}
where $\mathrm{FE}$ represents the feature extraction network composed of depth residual network and LSTM, $\mathrm{CAT}(\cdot)$ represents dimension splicing operation, $\mathrm{CLA}$ represents blood pressure classifier, REG represents blood pressure calculator, $\mathrm{R}_{\mathrm{cla}}$ represents the output features of classifier, $\mathrm{R}_{\mathrm{reg}}$ represents the output results of classifier and calculator respectively, and  $\mathrm{SOFTMAX}(\cdot)$ represents softmax operation.
We use the cross entropy loss function to train the classifier and MAE loss function to train the calculator. The two loss functions are combined to train the proposed model, which are formulated as follows:
\begin{equation}
\mathrm{Loss}_{\mathrm{cla}}=\frac{1}{N} \sum_{i}-\left[q_{i} \cdot \log \left(p_{i}\right)+\left(1-q_{i}\right) \cdot \log \left(1-p_{i}\right)\right]
\end{equation}
\begin{equation}
\mathrm{Loss}_{\mathrm{reg}}=\frac{\sum_{\mathrm{i}=1}^{N}|f_{i}-h_{i}|}{N}
\end{equation}
\begin{equation}
\mathrm{Loss} = \mathrm{Loss}_{\mathrm{cla}} +  \mathrm{Loss}_{\mathrm{reg}}
\end{equation}
where $p_{i}$ represents the result of classifier, $q_{i}$ represents the category label of blood pressure truth value, $f_{i}$ represents the result of calculator, $h_{i}$ represents the truth calue of blood pressure.
The final output result R needs to be calculated with the classifier result as the reference value and the deviation calibration of the calculator.
\begin{equation}
\mathrm{R}=\alpha \cdot \mathrm{STA}\left(\mathrm{R}_{\mathrm{cla}}\right)+ \beta \cdot \mathrm{R}_{\mathrm{reg}}
\end{equation}
where $\mathrm{{STA}}$ represents the reference value of each blood pressure interval and the deviation weight, and $\alpha$ represents the weight of the calculator results, $\beta$ represents the weight of the classifier results.

\subsection{Oversampling training strategy}\par
In order to deal with unabalance distribution of data in different blood pressure intervals in datasets, an oversampling training strategy is proposed to deal with the unbalance distribution of blood pressures. First, according to the distribution and level of blood pressure, different grouping strategies are designed for systolic and diastolic pressure data, which is then divided into four big groups based on the value of blood pressure $\left\{\mathrm{G}_{1}, \mathrm{G}_{2}, \mathrm{G}_{3}, \mathrm{G}_{4}\right\}$. In the five-fold cross validation experiment, each big group was divided into five small groups equally again, four of them are as training set and the other one is as the validation set. For each batch, the ratio of samples from each large group is 1:1:1:1. Data is extracted from each group in order. If the quantity is insufficient, the sample will be selected from the beginning. The sampling strategy to construct training dataset group ST(d,c) and validation dataset group SV(d,c) can be respectively computed as:

\begin{equation}
\mathrm{ST}(d, c)=\bigcup_{d=1}^{4}\left(G_{d c}\right)
\end{equation}

\begin{equation}
\mathrm{SV}(d, c)=\bigcup_{d=1}^{4}\left(\bigcup_{d=1}^{c} G_{d c}+\bigcup_{d=c+1}^{5} G_{d c}\right)
\end{equation}
where $G_{dc}$ represents a small dataset group, and $d$ and $c$ indicates the times of cross-validation and the index of samll group in each big group.

\subsection{Experiments}
subsection{Dataset}
The performance of the proposed network is evaluated on a public dataset and a private dataset. MMSE-HR~\cite{35Tulyakov2016} is a public non-contact heart rate and blood pressure estimation database composed of 102 face videos from 40 subjects and recorded at 25 frames per second (fps). A physiology data acquisition system is used to collect the average HR and BP values.\par
\begin{figure}[h!]
	\centering
\subfigure[The distributions of the systolic blood pressure values in ASPD dataset]
	{	\includegraphics[width=0.7\linewidth]{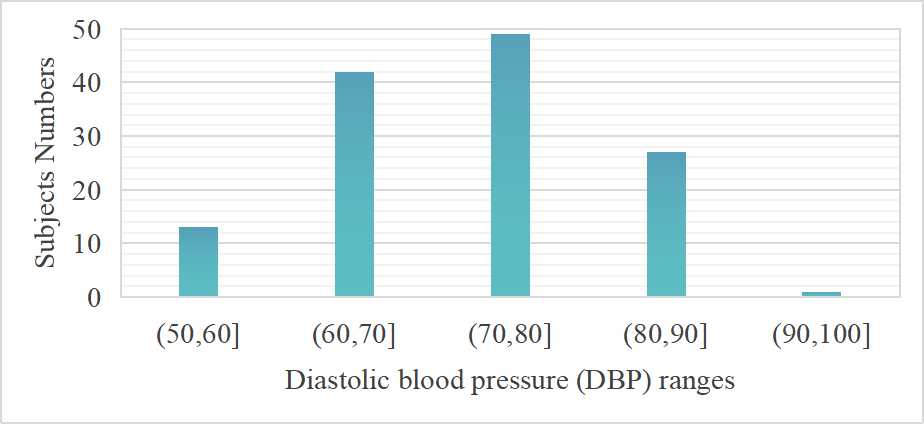}
	}\\
\subfigure[The distributions of the diastolic blood pressure values in ASPD dataset]
	{	\includegraphics[width=0.7\linewidth]{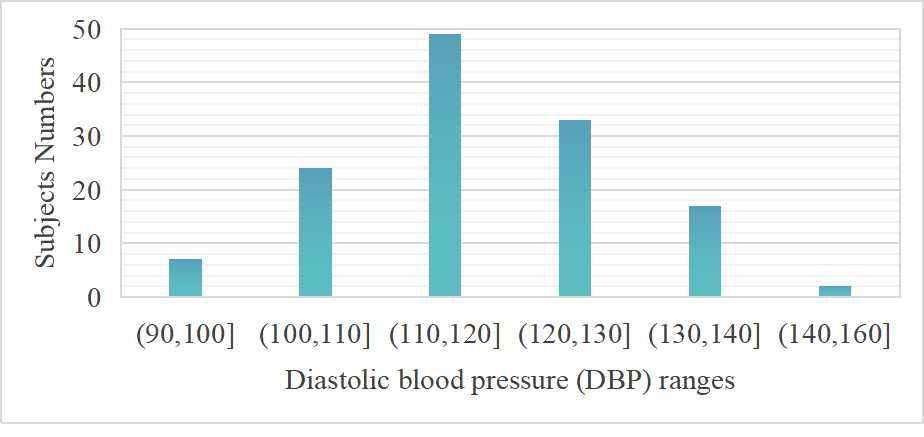}
	}
	  \caption{The distributions of the ground-truth blood pressure values in ASPD dataset. (a) The systolic blood pressures. (b) The diastolic blood pressures.}
	  \label{fig3}
\end{figure}
\begin{figure}[ht]
	\centering
		\includegraphics[width=.5\linewidth]{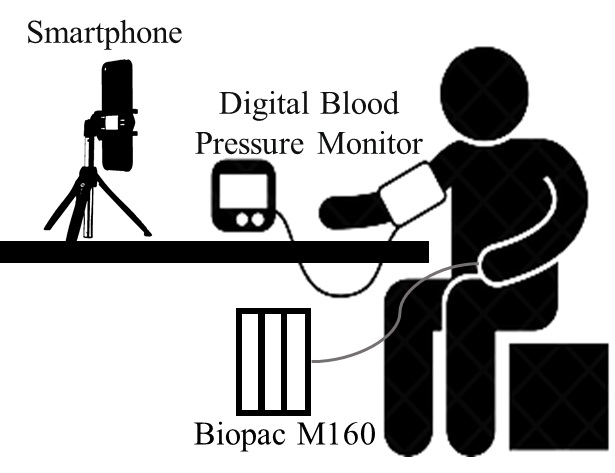}
	  \caption{Devices and setup used in collecting ASPD.}
	  \label{fig5}
\end{figure}

A private database for non-contact blood pressure estimation is created, named as Advanced Sensor Physiological Dataset (ASPD). In this dataset, the physiological data and corresponding face videos of 124 people are collected by using a multi-channel physiological signal acquisition system (Biopac M160), a blood pressure monitor (OMRON HEM-1020), and a mobile phone camera (Huawei Mate30), respectively. The face video for each subject is collected for about 1 minute. The frame rate of the video is 30fps, and the resolution is 1920×1080. The distributions of systolic and diastolic blood pressures of ASPD dataset are shown in Figure \ref{fig3}. Moreover, the devices and setup is illuminated in Figure~\ref{fig5}.\par

Three metrics were used to evaluate the performance of the network including the standard deviation (SD), the root mean square error (RMSE), and the mean absolute error (MAE).\par
In ASPD cross validation, all face videos are sampled to 30fps. Training is done with Nvidia-v100 and Pytorch. The authors design an independent oversampling learning strategy for systolic and diastolic blood pressure, and conduct independent cross validation and cross dataset testing.  Firstly, a five-fold cross validation is performed on ASPD dataset, and comparative experiments are conducted to verify various key module parameters of the network and the optimal combination scheme. After that, ASPD is used to train and test cross datasets on MMSE-HR. Finally, a five-fold cross-validation is performed on ASPD, and ablation experiments were done on the oversampling training scheme to verify its effectiveness.\par

\subsection{Ablation study}
\subsubsection{The effectiveness of spatial-temporal slicer}
\begin{table}[h!]
\centering
\caption{\centering{Performance comparison of different network.}}\label{tb4}
\setlength{\tabcolsep}{1mm}{
\begin{tabular}{ccccc}
\toprule
 & Method & SD$(\mathrm{mmHg})$ & RMSE $(\mathrm{mmHg})$ & MAE$(\mathrm{mmHg})$ \\ 
\midrule
 \multirow{4}*{SBP} & Without LSTM & $10.49$ & $17.85$ & $15.78$ \\
 & LSTM-90 & $10.06$ & $10.33$ & $8.44$ \\
 & LSTM-150 & \textcolor{red}{9.81} & \textcolor{red}{9.94} & \textcolor{red}{8.07}\\
 & LSTM-225 & $10.09$ & $10.16$ & $8.32$ \\

\midrule
\multirow{4}*{DBP}  
 & Without LSTM & $9.28$ & $17.06$ & $15.04$ \\
 & LSTM-90 & $8.35$ & $9.12$ & $7.31$ \\
 & LSTM-150 & \textcolor{red}{8.28} & \textcolor{red}{8.45} & \textcolor{red}{6.78}\\
 & LSTM-225 & $8.33$ & $9.05$ & $7.33$ \\
\bottomrule
\end{tabular}}
\end{table}
In order to verify the effectiveness of the combination of the spatiotemporal slicer and LSTM, we use the oversampling training scheme to train systolic and diastolic blood pressure, respectively. At the same time, they all use ResNet18 as the backbone network and YUVT as the color space. It is predicted that the specific length of the spatiotemporal slicer would greatly affect the final training results, and the optimal scheme of spatiotemporal slicing can be found. Clip length is tested for 90,150,225 and 450. \par
From Table~\ref{tb4}, spatiotemporal slicer with LSTM can help the network obtain the detailed characteristics of physiological information more effectively, and the slice length of 150 frames can maximize the effect of this network combination. For example, as shown in Table~\ref{tb4}, MAE of systolic blood pressure and diastolic blood pressure reached 8.07mmHg and 6.78mmHg, respectively. Compared with other slice length schemes, MAE of systolic blood pressure and diastolic blood pressure increases by 0.37mmHg and 0.55mmHg, respectively, which is much better than the network scheme without temporal and spatial slices (15.78 mmHg and 15.04 mmHg, respectively). Then observe SD and RMSE. In practical measurement, the number of observations n is always limited, and the actual value can only be replaced by the most reliable (optimal) value. RMSE is susceptible to a group of significant or minor errors in measurement, so it can well reflect measurement precision. It can be seen from the table that RMSE and SD of the 150 slice length scheme are still the best, which shows that this scheme not only improves accuracy but also enhances stability.\par
\subsubsection{The effectiveness of oversampling strategy(OSS)}

\begin{table}[h!]
\centering
\caption{\centering{Performance comparison of different networks.}}\label{tb5}
\setlength{\tabcolsep}{2mm}{
\begin{tabular}{ccccccccc}
\toprule
 & Method & SD$(\mathrm{mmHg})$ & RMSE $(\mathrm{mmHg})$ & MAE$(\mathrm{mmHg})$ \\ 
\midrule
 \multirow{2}*{SBP} & With OSS & \textcolor{red}{9.81} & \textcolor{red}{9.94} & \textcolor{red}{8.07}\\
& Without OSS & $10.12$ & $10.14$ & $8.27$ \\
\midrule
\multirow{2}*{DBP} & With OSS & \textcolor{red}{8.28} & \textcolor{red}{8.45} & \textcolor{red}{6.78} \\
& Without OSS & $8.35$ & $8.6$ & $6.92$ \\
\bottomrule
\end{tabular}}
\end{table}\par
Then the effectiveness of the oversampling training scheme is tested, with the standard data sampling training scheme as a comparison (in each epoch, training samples were imported into the network for training). The above results show that the combination of spatiotemporal slice and LSTM is the best when clip length is 150. Therefore, in this test, LSTM-150 is used as the primary network structure. Table \ref{tb5} shows that the training scheme enabled the same network model with more vital fitting ability and robustness. MAE of systolic blood pressure and diastolic blood pressure decreases from 8.27mmHg and 6.92mmHg to 8.07mmHg and 6.78mmHg. RMSE also decreases by 0.2mmHg and 0.15mmHg, respectively. This shows that for the blood pressure task with few training samples, appropriate oversampling training scheme is helpful to further enhance the network potential.\par

\subsection{Cross-dataset testing}
\begin{figure}[h]
	\centering
	\subfigure[The Bland-Altman plots of the estimated systolic blood pressure values]
	{\includegraphics[width=0.6\linewidth]{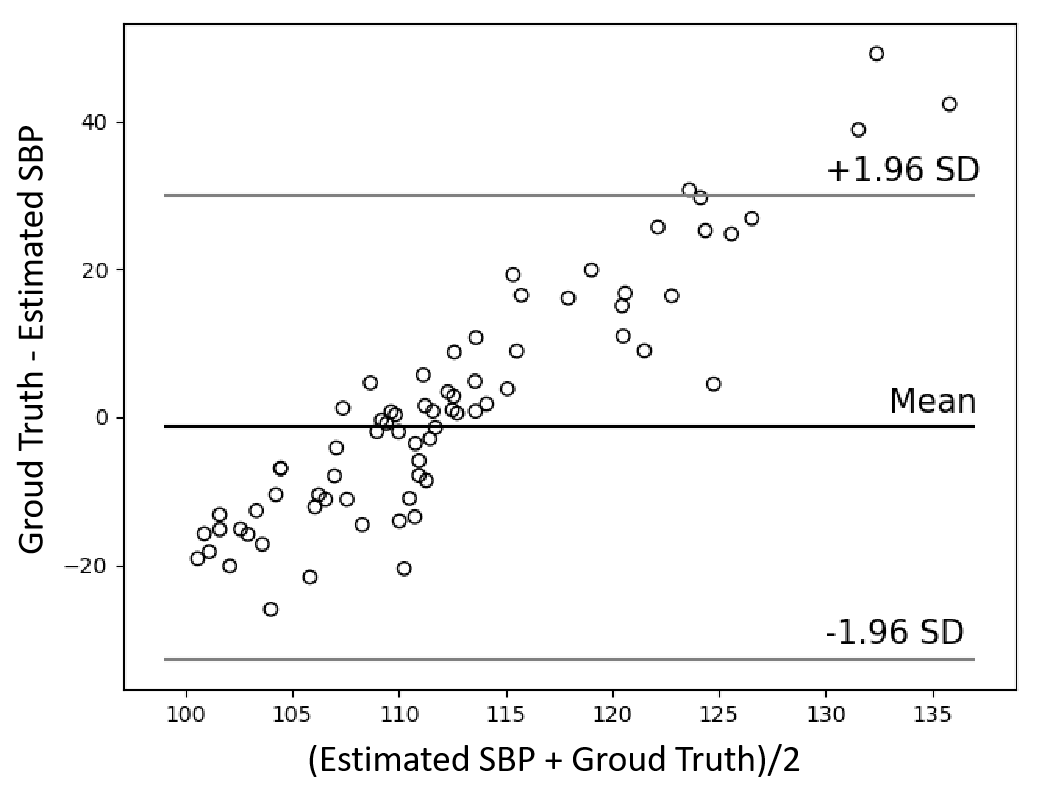}
	}\\
	\subfigure[The Bland-Altman plots of the estimated diastolic blood pressure values]
	{\includegraphics[width=0.6\linewidth]{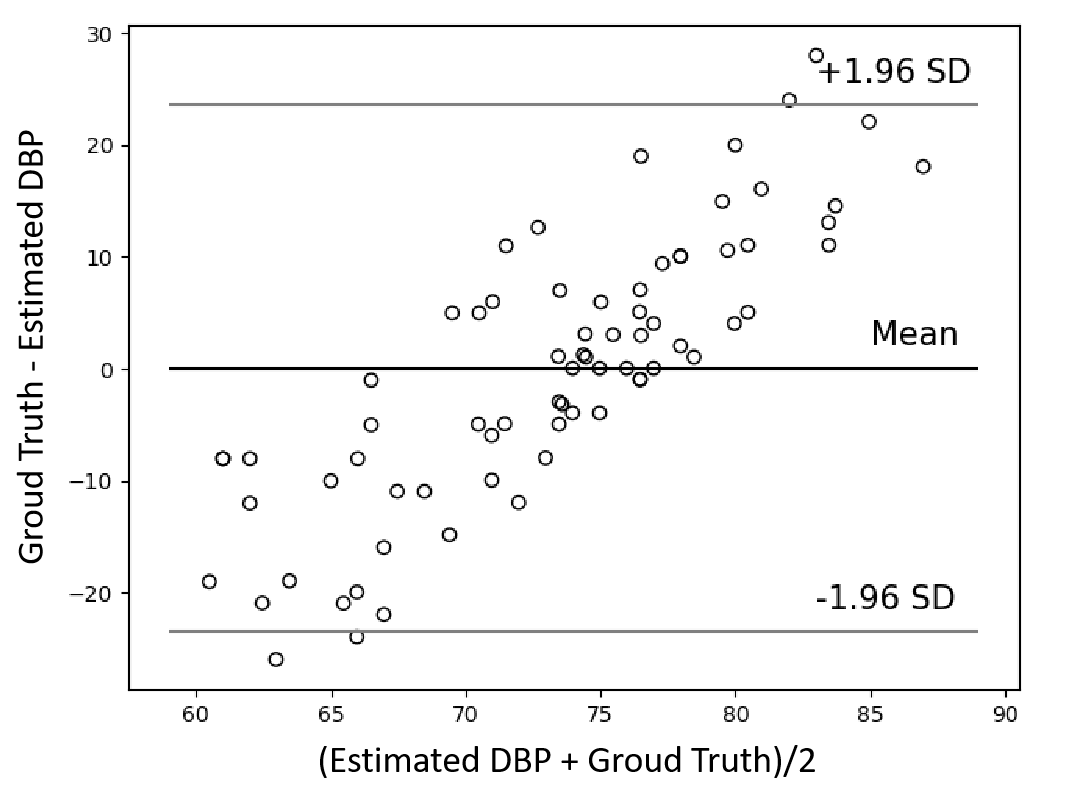}
	}\\
\caption{The Bland-Altman plots of the estimated (a) systolic blood pressure and (b) diastolic blood pressures comparing with their ground truths.
}
	\label{fig6}
\end{figure}

A recent study provide a solution for blood pressure estimation based on rPPG. It used thousands of data for training, and 50 samples for test~\cite{43SF2021}. For our BPM-Net, 125  data samples are used for training and MMSE-HR data are used for testing. The sample ratio tends to be close to 1:1. After three fine-tuning of the model, our BPM-Net obtained the best performance for the test is that the MAE of systolic blood pressure is 13.6mmHg and the MAE of diastolic blood pressure is 10.3mmHg. The comparative results are shown in Table\ref{tb3}. In this study, we reproduced the NCBP~\cite{44Rong2021} algorithm, trained it with ASPD and tested it on MMSE-HR. Besides the results of NIBPP~\cite{43SF2021} on its test data set (about 20 people) are also placed in the table for comparison. From the results, it can be seen that, our method get better performance than the recent study.  In order to verify the effectiveness of BPM-Net, the Bland-Altman plots as showed in Figure~\ref{fig6} show a good consistency between the estimated blood pressure and the ground truth.

\begin{center}
\begin{table}[h!]
\centering
\caption{\centering{Results of BP measurements}}\label{tb3}
\setlength{\tabcolsep}{2.5mm}{
\begin{tabular}{ccccc}
\toprule
& Method & SD($\mathrm{mmHg}$) & RMSE $(\mathrm{mmHg})$ & MAE($\mathrm{mmHg}$) \\ 
\midrule
\multirow{3}*{SBP} & NIBPP~\cite{43SF2021} & $-$ & $-$ & $13.6$ \\
~ & NCBP~\cite{44Rong2021} & $19.81$ & $22.43$ & $17.52$ \\
~ & LSTM-90 & $17.02$ & $17.35$ & $13.42$ \\
~ & LSTM-150 & \textcolor{red}{16.02} & \textcolor{red}{16.55} & \textcolor{red}{12.35} \\
~ & LSTM-225 & $16.98$ & $17.12$ & $13.15$ \\
\midrule
\multirow{3}*{DBP} & NIBPP~\cite{43SF2021} & $-$ & $-$ & $10.3$ \\
~ & NCBP~\cite{44Rong2021} & $15.21$ & $15.23$ & $12.13$ \\
~ & LSTM-90 & $13.34$ & $13.44$ & $10.41$ \\
~ & LSTM-150 & \textcolor{red}{11.98} & \textcolor{red}{12.22} & \textcolor{red}{9.54} \\
~ & LSTM-225 & $12.87$ & $13.22$ & $10.33$ \\
\bottomrule
\end{tabular}}
\end{table}
\end{center}

\section{Conclusion}
This paper presents the first non-contact end-to-end blood pressure measurement network, which only uses face video information to quickly calculate diastolic and systolic blood pressure in 15 seconds. Tests have been run on the public dataset MMSE-HR and the self-collected data set ASPD, and the results show that the proposed BPM-Net can effectively and efficiently measure blood pressure.

\section{Acknowledgments}
This work is supported by the Sichuan Science and Technology Program under Grant 2022YFS0032 to Xiujuan Zheng, the Shaanxi Provincial Natural Science Basic Research Program under Grant 2021JQ-455 to Bin Li. The authors would like to thank the engineers of Xi'an Singularity Fusion Information Technology Co. Ltd for their supports of data collection and experimental procedures.


\section*{Declaration of competing interest}
The authors declare that they have no known competing financial interests or personal relationships that could have appeared to influence the work reported in this paper.\par


\bibliography{mybibfile.bib}
\bibliographystyle{ieeetr}
\end{document}